\documentclass[aps,prl,twocolumn,superscriptaddress,showpacs,floatfix]{revtex4}

\usepackage{bm}
\usepackage{amsmath}
\usepackage{amssymb}
\usepackage{graphicx}
\usepackage{soul}
\usepackage{color}
\usepackage{xcolor}\definecolor{ao}{rgb}{0.0,0.0,1.0}

\definecolor{br}{rgb}{1.0, 0.22, 0.0}

\usepackage{times}

\input{epsf}
\begin{document}
\title{
Efficiency bounds on thermoelectric transport in magnetic fields: \\
The role of  inelastic processes \\
}

\author{Kaoru Yamamoto}
\affiliation{Department of Physics, The University of Tokyo, Komaba 4-6-1, Meguro, Tokyo 153-8505, Japan }
\email{kaoru3@iis.u-tokyo.ac.jp}

\author{Ora Entin-Wohlman}
\affiliation{Raymond and Beverly Sackler School of Physics and Astronomy, Tel Aviv University, Tel Aviv 69978, Israel}
\affiliation{Physics Department, Ben Gurion University, Beer Sheva 84105, Israel}

\email{oraentin@bgu.ac.il}

\author{Amnon Aharony}
\affiliation{Raymond and Beverly Sackler School of Physics and Astronomy, Tel Aviv University, Tel Aviv 69978, Israel}
\affiliation{Physics Department, Ben Gurion University, Beer Sheva 84105, Israel}

\author{Naomichi Hatano}
\affiliation{Institute of Industrial Science,  The University of Tokyo, Komaba 4-6-1, Meguro, Tokyo 153-8505, Japan}
\email{hatano@iis.u-tokyo.ac.jp}

\date{\today}

\begin{abstract}
We examine the efficiency of an effective two-terminal thermoelectric device under  broken time-reversal symmetry. The setup is derived  from a three-terminal thermoelectric device comprising a thermal terminal and two electronic contacts, under a magnetic field.  
We find that breaking time-reversal symmetry
in the presence of the inelastic electron-phonon processes can
significantly enhance the figure of merit for delivering electric
power by supplying heat from a phonon bath, beyond  the one for
producing the electric power by investing thermal power from the
electronic heat current. The efficiency of such a device is
bounded by the non-negativity of the entropy production of the
original three-terminal junction. The efficiency at maximal power can
be quite close to the Carnot efficiency, 
but then the electric power vanishes. 
\end{abstract}


\maketitle

Achieving high efficiencies 
in thermoelectric nanodevices is one of the main goals of contemporary research on nano-structural materials and setups  \cite{Dresselhaus,Vineis,Shakouri}.
It is well known that the efficiency (the ratio of the output power to the invested one) of
a heat engine operating  between two reservoirs  is bounded by  the Carnot efficiency.
This extreme efficiency is reached in a quasi-static process  which lasts for an infinite time and therefore the extracted power vanishes.
The maximal efficiency  that can be achieved by a heat engine delivering a finite electric power in a two-terminal  geometry  is usually expressed in terms of a single material- and setup-dependent   parameter,  called the figure of merit, $\zeta$.   The larger the $\zeta$, the higher   the maximal attainable efficiency can be.
The possibility to significantly enhance  $\zeta$ 
of a nanostructure,  and even to reach the limit of reversibility  by exploiting transmission resonances, has been extensively discussed  \cite{Mahan,Humphrey1,Humphrey2};  the  maximal efficiency attained at finite output powers \cite{Whitney} has been explored along similar lines.
 A recent theoretical endeavor  introduced    the concept of stochastic efficiencies, pertaining to  strong fluctuations in the efficiency of small systems
(see, e.g., Refs.~\onlinecite{Polettini,
Proesmans,Esposito,Jiang}).
The bound on this efficiency depends on the protocol by which  the long-time limit is reached.

Another promising route of research is the
multi-terminal nanoscale heat engines \cite{Mazza,BS}, e.g., those built on coupled quantum dots \cite{Molenkamp,Molenkamp1}.
In contrast to  the two-terminal geometry, where heat and charge
are carried by the same particles, it is possible in
the three-terminal setups  to spatially separate the heat reservoir from the current circuit \cite{Molenkamp,Benenti1}, thus improving  
the  functionality  of the device. Here we explore the possibility to  supply heat from a nonelectronic source to produce electric power,  and examine the efficiency 
as compared to the one of a conventional electronic two-terminal device. 

The theoretical attempts to improve $\zeta$ \cite{Mahan,Humphrey1,Humphrey2}  mostly
pertain to
setups which are time-reversal symmetric.
This means that the matrix relating the fluxes (i.e.,  currents of particles and heat currents)
to the thermodynamic driving forces (electric voltages and temperature differences) is symmetric \cite{Onsager,Casimir}. This matrix is composed of the Onsager coefficients, which
 in the {\em linear-response regime} are independent of the driving forces or the fluxes.
An intriguing  paper by Benenti {\it et al.}~\cite{Benenti}
raised the possibility of manipulating the performance of electronic thermoelectric devices by breaking  time-reversal symmetry.
In fact,  invoking solely the Onsager
reciprocity relations \cite{Onsager,Casimir,Mathews,Ludovico} and the non-negativeness of the entropy production,  Ref.~\onlinecite{Benenti} obtained the perplexing possibility of
a two-terminal (2T) device operating at the Carnot efficiency while yielding a finite power.
This result has been investigated in the literature of not only the quantum thermoelectricity,  but also classical and quantum heat engines \cite{Brandner2015PRX, Proesmans2015PRL, Shiraishi2016arXiv}.

How can this claim be scrutinized?
Time-reversal symmetry can be broken by magnetic fields.
However, 
the
linear-response Onsager coefficients of an electronic 2T device without interactions, i.e., with only elastic scattering, 
must be even functions of the magnetic field \cite{Buttiker}.
Thus, a prerequisite for this investigation
is to find a realistic situation in which the asymmetry of the Onsager coefficients can be controlled.
Here we respond to this challenge by adding inelastic processes.
Our 
system allows for the inspection of the effect of a broken time-reversal symmetry in conjunction with inelastic processes,   on the power-harvesting efficiency.

\begin{figure}
\vspace{0.cm} 
\includegraphics[width=4.5cm]{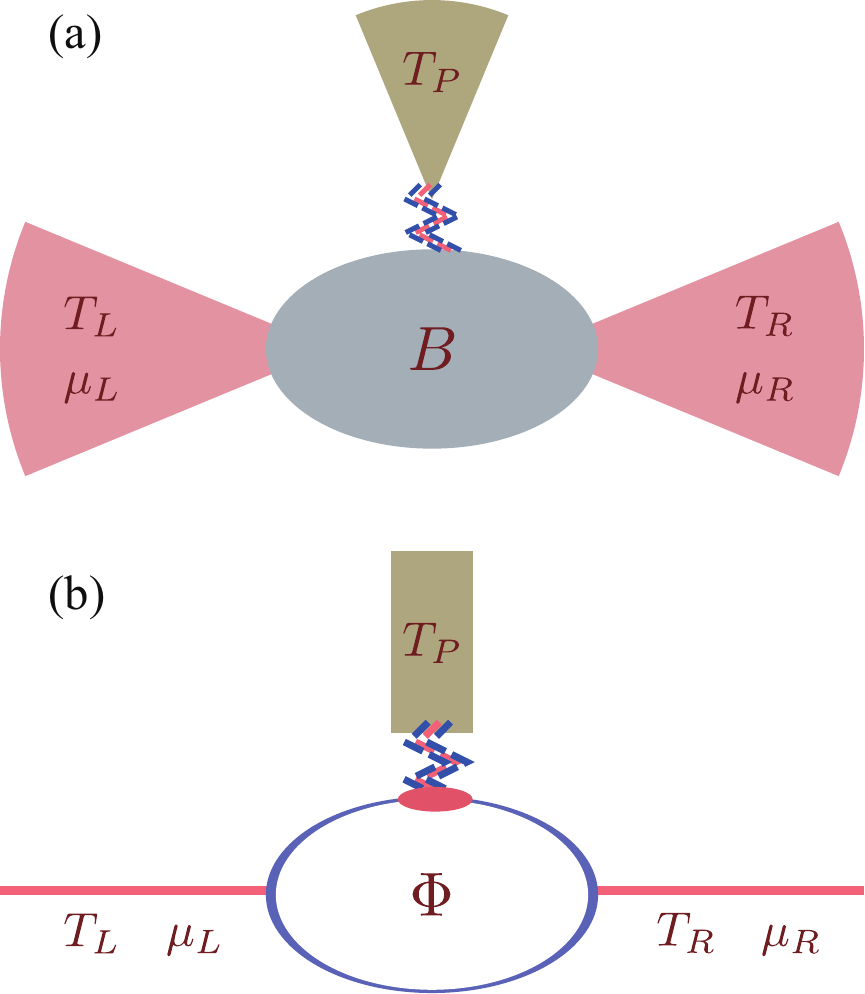}
\caption{(color online)
Sketch (a) and  simplified model (b) of a  thermoelectric device, comprising two electronic terminals (held at chemical potentials $\mu^{}_{L}$ and $\mu^{}_{R}$, and at temperatures $T^{}_{L}$ and $T^{}_{R}$), and a thermal terminal held at a temperature $T_{P}$, with which the electrons exchange energy.  The nanostructure [central (gray) disk],   placed in a magnetic field $B$ as in (a),  is modeled  by an Aharonov-Bohm ring threaded by a magnetic flux $\Phi$ as in (b). The electrons exchange energy with vibrational modes on  a dot placed on the upper arm. 
}
\label{fig1}
\end{figure}

An effective two-terminal (E2T)
setup  can be constructed from  an all-electronic three-terminal (3T) device
\cite{Saito,Brandner,Balachandran}  (or even from a multi-terminal
electronic junction \cite{Brandner1}). 
This proposal is based on the concept of probe terminals  \cite{Buttiker1}, whose temperatures and chemical potentials can be adjusted so as to cancel out the currents between those terminals and the device. In this way the multi-terminal setup is reduced to an E2T one. 
Though
breaking time-reversal symmetry can enhance the
efficiency of the E2T device,   the unitarity of the 3T scattering
matrix imposes strong bounds on the efficiency of the E2T, which
are much lower than those of Ref.~\onlinecite{Benenti}
\cite{Brandner, Balachandran}.
Adding more probe
terminals can increase these bounds and the resulting efficiency,
but these still remain below those of
Ref.~\onlinecite{Benenti} \cite{Brandner1}.

The matrix of the Onsager coefficients can be asymmetric when time-reversal symmetry is broken in the presence of inelastic interactions \cite{Sanchez},
or 
electron-phonon ones as considered here  \cite{PRB12}.  These
 interactions give rise to inelastic processes between the conduction electrons and the phonons, and those in turn imply the existence of  thermal baths attached to the junction that
exchange energy  (but not particles) with the electronic system. Figure  \ref{fig1}(a) illustrates a setup of this type \cite{PRB12}: a nanostructure is attached to two electronic baths       held at chemical potentials
$\mu_{L}$ and $\mu_{R}$, and at temperatures $T_{L}$ and $T_{R}$, and is coupled to a third, thermal bath, held at another temperature $T_{P}$.
There are three independent currents flowing in this device, namely, the electric one  between the electronic baths, the electronic heat current between them, and the phonon energy current.
This setup becomes an E2T 
one with asymmetric Onsager coefficients when e.g., the driving forces are chosen so that the electronic heat current is 
blocked.
In this way 
the device investigated in Ref.~\onlinecite{Benenti} is realized.

We find that while the entropy production of the 3T 
junction can vanish, that of the E2T 
junction cannot, as long as the magnetic field $B$ is nonzero \cite{CommentB}.
In other words,
the E2T device is dissipative as long as $B\neq 0$, and its 
efficiency never reaches the Carnot bound.  Though the argument of Ref. \onlinecite{Benenti} is correct for the purely 2T setup, 
the upper bound on the efficiency is reduced. However, this upper bound increases with the strength of the electron-phonon interaction, and can exceed  that found for the 3T all-electronic junctions \cite{Brandner}.

The entropy production of the nanostructure in Fig. \ref{fig1}(a) is
\begin{align}
\dot{S}=\frac{\langle \dot{E}^{}_{P}\rangle}{T_{P}^{}}+\frac{\langle \dot{E}^{}_{L}\rangle -\mu^{}_{L}\langle \dot{N}^{}_{L}\rangle}{T^{}_{L}}
+\frac{\langle \dot{E}^{}_{R}\rangle -\mu^{}_{R}\langle \dot{N}^{}_{R}\rangle}{T^{}_{R}}\ ,
\end{align}
where  $-\langle\dot {E}_{L(R,P)}\rangle$ is the energy current out of the left electronic
bath (right electronic bath, phonon bath) and $-\langle \dot{N}_{L(R)}\rangle $ is the  particle current  out of
the left (right) electronic bath. The particle number conservation,  $\langle \dot{N}_{L}+\dot{N}_{R}\rangle =0$, uniquely
identifies
the charge current, $J_{L}=-e\langle\dot{N}_{L}\rangle =e\langle\dot{N}_{R}\rangle$ ($e$ is the unit charge). In contrast,
energy conservation, $\langle \dot{E}_{P}+\dot{E}_{L}+\dot{E}_{R}\rangle =0$,
does not yield  unique identifications of the energy currents \cite{Yamamoto2015arXiv}.
Choosing $T_{R}$  as the reference,  using the electronic heat current emerging from the left electronic bath \cite{Yamamoto2015PRE}, we have
\begin{align}
J^{Q}_{L}=-\langle \dot{E}^{}_{L}\rangle+\mu^{}_{L}\langle \dot{N}^{}_{L}\rangle\ ,
\end{align}
and the heat current  between the thermal terminal and the electrons, $J^{Q}_{P}=-\langle \dot{E}_{P}^{}\rangle$,
the entropy production becomes
\begin{align}
T^{}_{R}\dot{S}=VJ^{}_{L}+\delta t^{}_{\rm el}J^{Q}_{L}+\delta t^{}_{\rm e-p}J^{Q}_{P}\ .
\label{ent1}
\end{align}
The driving forces are the voltage drop,
$V=(\mu^{}_{L}-\mu^{}_{R})/e $, the (dimensionless) difference in the electronic temperatures,  $\delta t^{}_{\rm el}=1-T^{}_{R}/T^{}_{L}$, and the temperature difference between the phonon bath and the reference, $\delta t^{}_{\rm e-p}=
1-T^{}_{R}/T^{}_{P}$.

We consider the model displayed in Fig.~\ref{fig1}(b),  in which the nanostructure is an Aharonov-Bohm ring, threaded by  a magnetic flux $\Phi$.
The ring is attached to two electronic reservoirs and carries a quantum dot on one of its arms; when the electrons are on the quantum dot they interact  with the vibrational modes there. The latter are  tightly coupled to a bath   of phonons which fixes their population  \cite{PRB10}.
In our simplified model,  the dot is replaced by a single localized  electronic level of energy $\epsilon^{}_{0}$ and its vibrational modes are  Einstein phonons of frequency $\omega^{}_{0}$; the
electronic leads are assumed to contain   free electron gases.
With $\gamma$ denoting the coupling energy of an electron with the vibrational modes, whose creation (annihilation) operators are $b^{\dagger}$ ($b$),
and the creation (annihilation) operators of the electron  on the localized level by $c^{\dagger}_{0}$ ($c^{}_{0}$),
the Hamiltonian (using $\hbar =1$) is
\begin{widetext}
\begin{align}
{\cal H}&=[\epsilon^{}_{0}+\gamma (b^{\dagger}_{}+b)]c^{\dagger}_{0}c^{}_{0}+\omega^{}_{0}\Big(b^{\dagger}b+\frac{1}{2}\Big)
+\sum_{k}\epsilon^{}_{k}c^{\dagger}_{k}c^{}_{k}+\sum_{p}\epsilon^{}_{p}c^{\dagger}_{p}c^{}_{p}\nonumber\\
&+\sum_{k}(V^{}_{k}c^{\dagger}_{k}c^{}_{0}+{\rm H.c.})+
\sum_{p}(V^{}_{p}c^{\dagger}_{p}c^{}_{0}+{\rm H.c.})
+\sum_{k,p}(V^{}_{kp}e^{i\Phi}c^{\dagger}_{k}c^{}_{p}+{\rm H.c.})\ .
\label{ham}
\end{align}
\end{widetext}
The operators that create (annihilate) a conduction electron in the left (right) lead, of energy $\epsilon^{}_{k(p)}$,  are $c^{\dagger}_{k(p)}$ ($c^{}_{k(p)}$). The tunneling matrix elements between the localized level and the left  (right) electronic lead are denoted by $V^{}_{k}$ ($V^{}_{p}$). The lower arm of the ring in Fig.~\ref{fig1}(b) connects the two leads with the tunneling matrix element $V^{}_{kp}$; the magnetic flux $\Phi$ (in units of  $c/e$)  penetrating the ring is assigned to these elements. Since the magnetic field giving rise to the Aharonov-Bohm effect is usually small, one may neglect the tiny Zeeman effect on the spins of the conduction electrons.

The charge and heat currents flowing  through the ring junction are found within the Keldysh formalism 
\cite{Jauho,Commentin}. 
The detailed calculation, carried out to the second order in the coupling $\gamma$, 
is summarized in Ref.~\onlinecite{PRB12}.
The  charge current emerging from the left electronic lead is expressed  in terms of the electrons'   Green's functions, 
\begin{align}
J^{}_{L}=-e\langle\dot{N}_{L}^{}\rangle=-e\frac{d\langle \sum_{k}c^{\dagger}_{k}c^{}_{k}\rangle}{dt}=-e\int\frac{d\omega}{2\pi}{\cal J}^{}_{L}(\omega)\ ,
\label{JL}
\end{align}
where
\begin{align}
{\cal J}^{}_{L}(\omega )& =\sum_{k}V^{}_{k}[G^{}_{k0}(\omega )-G^{}_{0k}(\omega )]^{<}\nonumber\\
&+\sum_{k,p}V^{}_{kp}[e^{-i\Phi}G^{}_{kp}(\omega )-e^{i\Phi}
G^{}_{pk}(\omega)]^{<}\ .
\end{align}
Here, $G^{<}_{ab}(\omega )$ is the lesser Keldysh Green's function \cite{Jauho}, which is the Fourier transform of
$G^{<}_{ab}(t,t')=i\langle c^{\dagger}_{b}(t')c^{}_{a}(t)\rangle$, where $a$ and $b$ denote the relevant operators, [0, $k$, or $p$,  see Eq.~(\ref{ham})].
The energy current emerging from the left electronic lead is given by $-\langle\dot{E}_{L}\rangle =
d\langle \sum_{k}\epsilon_{k}^{}c^{\dagger}_{k}c^{}_{k}\rangle/dt$; it attains the same form as the charge current except for an extra factor $\omega$ in the integrand in Eq.~\eqref{JL} (and without the electron's charge $e$). The energy current of the thermal terminal  can be obtained from $\langle \dot{E}_{L}+\dot{E}_{R}+\dot{E}_{P}\rangle =0$.

Once the Green's functions are determined and inserted into the expressions for the currents, one expands the latter  to linear order in the three driving forces, to obtain the Onsager coefficients.
The result can be presented in a matrix form,
\begin{align}
\left [\begin{array}{c}J^{}_{L} \\J^{Q}_{L} \\ J^{Q}_{P}\end{array}\right]={\cal M} \left [\begin{array}{c}V \\\delta t^{}_{\rm el}\\ \delta t^{}_{\rm e-p}\end{array}\right ]\ ,
\label{M}
\end{align}
where ${\cal M}$ is the 3$\times $3 matrix of the Onsager coefficients of the device; its elements depend on the choice of the driving forces \cite{Yamamoto2015arXiv}.
Reference \onlinecite{PRB12} presents the elements of the matrix ${\cal M}$ for an arbitrary ring junction (i.e., that does not possess any special spatial symmetries).
All off-diagonal elements of ${\cal M}$ contain terms odd in the magnetic flux (which obey the Onsager reciprocity relations).
Remarkably enough, {\em all} these odd terms arise from inelastic processes in which  the charge carriers exchange energy ($\omega_{0}$) with the vibrational modes.  All elements of ${\cal M}$  also contain terms even in the flux; these arise from  elastic as well as  inelastic processes of the transport electrons \cite{PRB12}. 
In the absence of the coupling of the electrons with the vibrational modes, the 
Onsager coefficients would be even in the flux 
and ${\cal M}$ would be  symmetric.

For a spatially-symmetric junction,  the matrix ${\cal M}$ is \cite{PRB12}
\begin{align}
{\cal M}=\left [\begin{array}{ccc}G\ \ &\  {\cal S}G&\ 0\\   G{\cal S}\  &\  \kappa^{}_{0}&\ -P(1+a)\\0\ &\ -P(1-a)&\ 2P\end{array}\right ]\ ,
\label{ms}
\end{align}
where $a=\tau^{}_{0}\sin\Phi$, which is odd in the magnetic field, 
and  $\tau_{0}$
is the  transmission   of the lower arm of the ring (assumed for simplicity to be energy independent).
In Eq. (\ref{ms}), $G$ is the electric conductance and ${\cal S}$ is the Seebeck coefficient; $\kappa_{0}$ is the ``bare" thermal conductance of the electrons, i.e., for $\gamma=0$, 
the heat conductance of the electrons is $\kappa_{0}-G{\cal S}^{2}$. These three coefficients, for the ring geometry of Fig.~\ref{fig1}(b), are functions of $\cos\Phi$ \cite{PRB12}. 
The other three elements in ${\cal M}$
are due  to inelastic processes (and vanish at zero temperature),  
\begin{align}
P=\omega^{2}_{0}\int\frac{d\omega}{\pi}
{\cal T}^{}_{\rm p}(\omega,\Phi)\ ,
\end{align}
where ${\cal T}_{\rm p}$ is the transmission  of the inelastic processes,
\begin{align}
{\cal T}^{}_{\rm p}(\omega ,\Phi)=\frac{\beta^{}_{R}}{e^{\beta^{}_{R}\omega^{}_{0}}-1}f(\omega^{}_{-})[1-f(\omega^{}_{+})]c(\omega,\Phi)\ .
\end{align}
Here,  $f(\omega)=(\exp[\beta^{}_{R}(\omega-\mu^{}_{R})]+1)^{-1}$ (recall that temperatures and chemical potentials are measured with respect to the right electronic lead), $\omega_{\pm}=\omega\pm\omega_{0}/2$,
\begin{align}
c(\omega,\Phi)&=\frac{\gamma^{2}}{4}\Gamma(\omega^{}_{-})\Gamma(\omega^{}_{+})|{\cal G}^{a}_{00}(\omega^{}_{-},\Phi){\cal G}^{a}_{00}(\omega^{}_{+},\Phi)|^{2}\ ,
\end{align}
$\Gamma$ represents the width of the resonance on the dot due to the coupling with the electronic leads, and ${\cal G}^{a}_{00}$ is the advanced Green's function there 
in the absence of the coupling with the vibrations; 
$2P\geq 0$ is the heat conductance of the phonons, proportional to $\gamma^{2}$,  the electron-phonon coupling squared \cite{PRB12}.

Inserting the explicit expressions for the currents [Eqs.~(\ref{M}) and (\ref{ms})] into Eq.~(\ref{ent1}), one finds that the entropy production of the  3T 
device is non-negative, $\dot{S} \geq 0$,  for
\begin{align}
\kappa^{}_{0}-G{\cal S}^{2}-P/2\geq 0\ ,\ \ {\rm i.e.,}\ \   2\kappa^{}_{0}/P\geq 1+\zeta\ ,
\label{con1}
\end{align}
where $\zeta=G{\cal S}^{2}/(\kappa^{}_{0}-G{\cal S}^{2})$ is the figure of merit of the conventional electronic 2T device [see e.g., Ref.  \onlinecite{Mahan}].

The electronic heat current $J_{L}^{Q}$ can be blocked 
by choosing 
\begin{align}
\delta t^{}_{\rm el}=-\frac{1}{\kappa^{}_{0}}[G{\cal S}V-P(1+a)\delta t_{\rm e-p}]\ .
\end{align}
The setup then
becomes an E2T,  
 in which electric power is produced  at the expense of  thermal power from the phonon bath. The matrix of the Onsager coefficients pertaining to this configuration is not symmetric, that is, the off-diagonal elements are not even  in the magnetic field \cite{comment},
\begin{align}
\left [\begin{array}{c}J^{}_{L}  \\ J^{Q}_{P}\end{array}\right]= \left [\begin{array}{cc}\frac{G}{1+\zeta}& \frac{PG{\cal S}}{\kappa^{}_{0}}(1+a)\\
\frac{PG{\cal S}}{\kappa^{}_{0}}(1-a)&\frac{P^{2}}{\kappa^{}_{0}}(\zeta^{}_{\rm max}+a^{2})\end{array}\right ]
\left [\begin{array}{c}V \\ \delta t^{}_{\rm e-p}\end{array}\right ]\ ,
\label{M2}
\end{align}
where 
$\zeta_{\rm max}=-1+2\kappa_{0}/P$ is the upper bound on $\zeta$, imposed by the condition (\ref{con1}).
The entropy production in the E2T 
setup described by Eq. (\ref{M2}), which is proportional to $J^{}_{L}V+J^{Q}_{P}\delta t^{}_{\rm e-p}$,   is non-negative
for
\begin{align}
 2\kappa^{}_{0}/P+a^{2}\geq 1+\zeta\ . \label{con2}
\end{align}
The equality in Eq.~\eqref{con2} is what Benenti \textit{et al.}~\cite{Benenti} used in discussing the possibility to achieve 
the Carnot efficiency at a finite power. However, taking  account of the 3T 
setup, which is the background of the E2T  device, 
implies that the stricter inequality \eqref{con1} must hold, and that the equality of Eq.~\eqref{con2} is not achievable for a nonzero $a=\tau_0\sin\Phi$.  This means that as long as the magnetic field \cite{CommentB} is finite, the entropy production of the E2T 
junction cannot reach the reversible limit.  Thus, 
the Carnot bound is not reached, although the symmetry of the E2T
Onsager coefficients is broken.

\begin{figure}
\vspace{0.cm} 
{\includegraphics[width=6.cm]{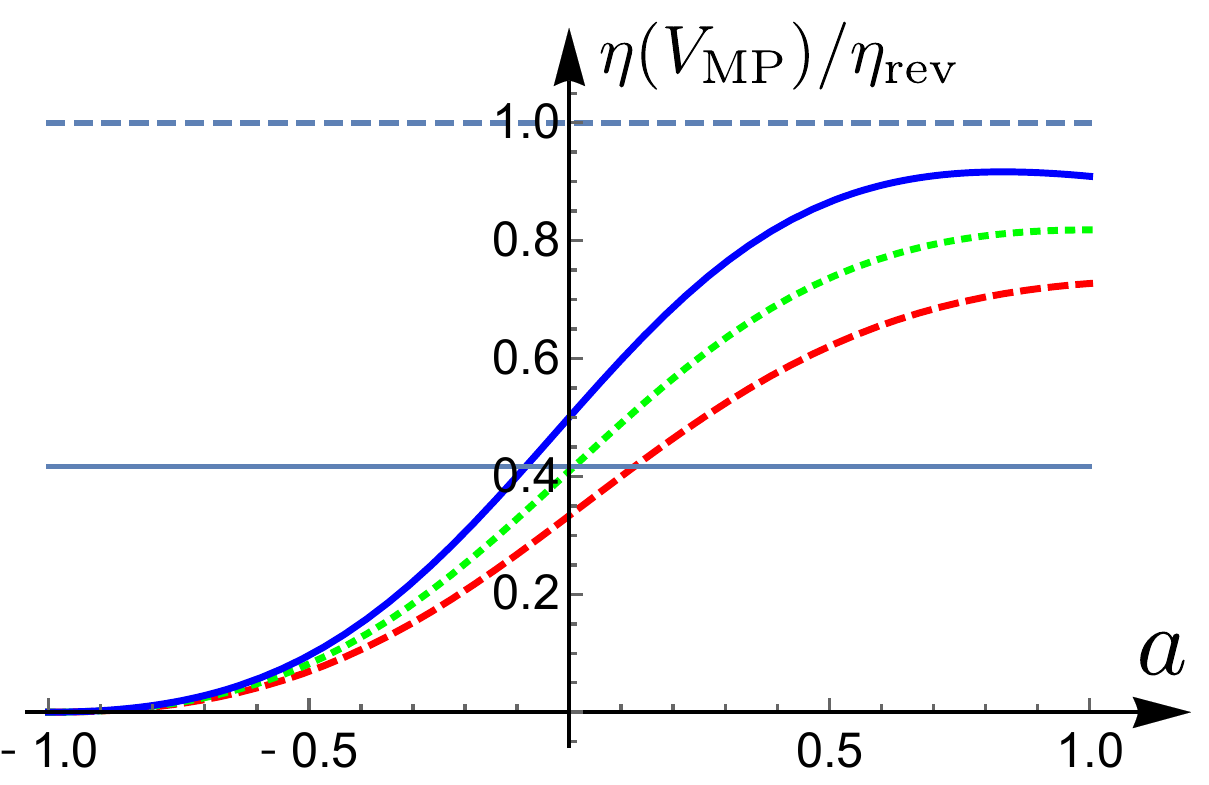}}
\caption{(color online)
The efficiency at maximal power of the E2T device (scaled by the Carnot efficiency $\eta_{\rm rev}$), Eq.~(\ref{etamp}),   as a function of the asymmetry parameter $a$, for $\zeta_{\rm max}=10$;  $\zeta=8$ [the dashed (red) curve], $9$ [the dotted (green) curve], and $10$ [the thick (blue) line]. The horizontal line is  $\eta(V_{\rm MP},\zeta=\zeta_{\rm max})/\eta_{\rm rev}$ of the  2T junction. 
}
\label{fig2}
\end{figure}

The efficiency for producing electric power at the expense of the phonons' energy current, i.e., $V\leq 0$ and $\delta t_{\rm e-p}\geq 0$,  is
\begin{align}
\eta=-VJ^{}_{L}/J^{Q}_{P}\ .
\label{efh}
\end{align}
The electric power $|VJ_{L}|$ is maximal at $V_{\rm MP}=-\delta t_{\rm e-p}(1+\zeta)(1+a)P{\cal S}/(2\kappa_{0})$, and then the efficiency at the maximal power is
\begin{align}
\eta(V^{}_{\rm MP})=\eta^{}_{\rm rev}
\frac{(1+a)^{2}\zeta/4}{\zeta^{}_{\rm max}+a^{2}-(1-a^{2})\zeta/2}\ ,
\label{etamp}
\end{align}
where $
\eta^{}_{\rm rev}=\delta t^{}_{\rm e-p}$ is the Carnot efficiency, 
i.e., $\eta(V_{\rm MP};\zeta=\zeta_{\rm max},a=0)=\eta_{\rm rev}/2$. Figure \ref{fig2}
displays $\eta(V^{}_{\rm MP})$     
of the E2T 
device as a function of 
the symmetry breaking parameter $a$ for several values of $\zeta\leq\zeta_{\rm max}=10$. The horizontal line shows the maximal electronic 2T 
efficiency at maximal power,
$\eta_{\rm max}(V_{\rm MP})/
\eta_{\rm rev}=\zeta_{\rm max}/(4+2\zeta_{\rm max})$.
Interestingly, all the graphs have a minimum,  $\eta(V_{\rm MP})=0$, at $a=-1$ 
(destructive interference on the Aharonov-Bohm ring), and a maximum at $a=1$ for $\zeta<\zeta_{\rm max}-1$ or at $a=(2\zeta_{\rm max}-\zeta)/(2+\zeta)$ for $\zeta>\zeta_{\rm max}-1$. This maximal value 
reaches the Carnot efficiency  when $\zeta=\zeta_{\rm max}\rightarrow\infty$, but  then the power vanishes.   
Remarkably enough,   $\eta(V_{\rm MP})$ 
of the  E2T 
is improved as the 
asymmetry in the Onsager coefficients is increased (except for a narrow region near $\zeta_{\rm max}$) and as $\zeta_{\rm max}=2\kappa_{0}/P-1$  increases; at large enough 
$\zeta$, the E2T 
device is more efficient than the 2T electronic one and is significantly higher than $\eta_{\rm max}(V_{\rm MP})/\eta_{\rm rev}=4/7$, found in Ref. \onlinecite{Brandner} for $a=1/7$.

\begin{figure}
\vspace{0.cm} 
{\includegraphics[width=6.5cm]{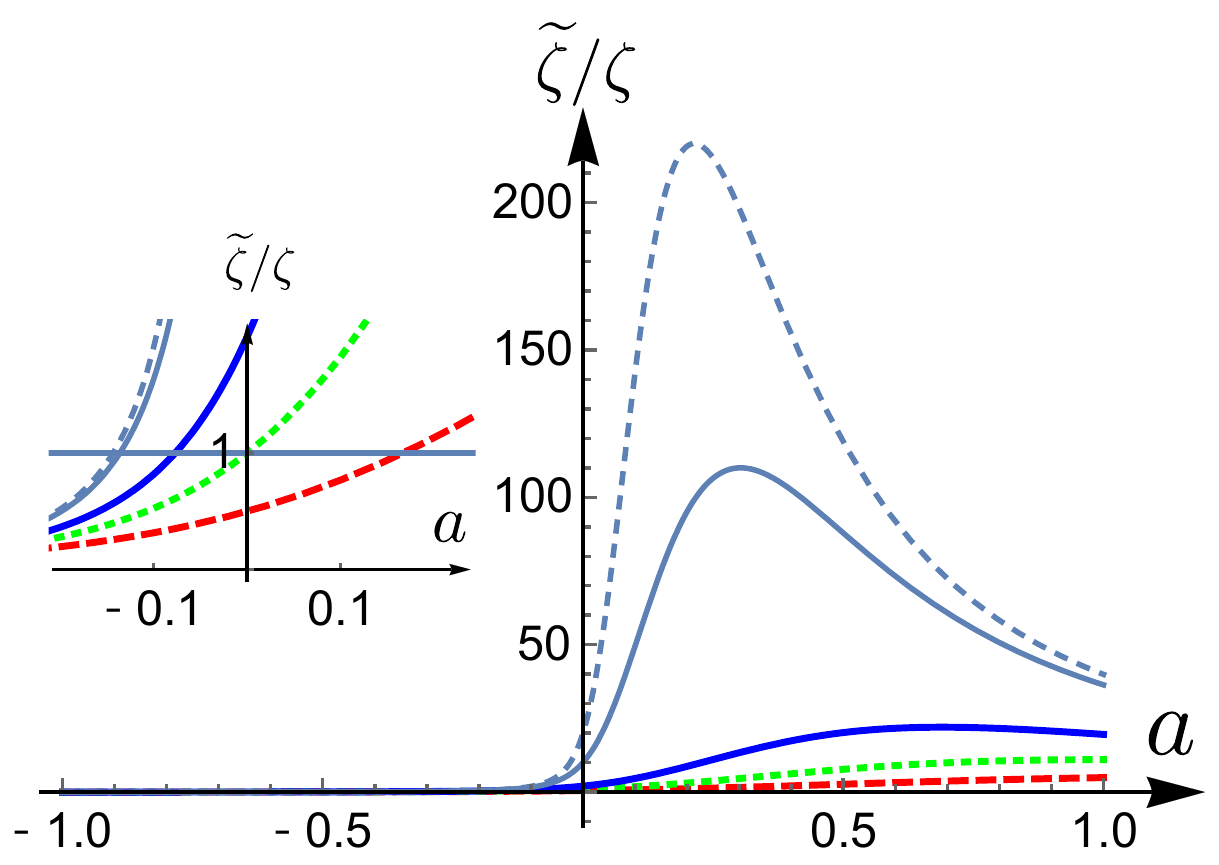}}
\caption{(color online)
The E2T  figure of merit $\widetilde{\zeta}$,  Eq. (\ref{newzeta}) (scaled by the  2T one,  $\zeta$),  as a function of  $a$, 
for $\zeta_{\rm max}=10$;  $\zeta=8$ [the dashed (red) curve], $9$ [the dotted (green) curve],  $9.5$ [the thick (blue) line],  $9.9$ [the thin (black) curve], and   $9.95$ [the thin  dashed  (black) curve]. Inset:
The range where 
$\widetilde{\zeta}/\zeta$ crosses 1.}
\label{fig3}
\end{figure}

Another quantity of interest is the maximal value of the efficiency Eq. (\ref{efh}),
reached when the voltage is
\begin{align}
\frac{V^{}_{\rm ME}}{\delta t^{}_{\rm e-p}}= \frac{P(\zeta^{}_{{\max}}+a^{2}_{})}{G{\cal S}(1-a)} \Big (\Big [1-\frac{(1-a^2)\zeta}{\zeta_{{\max}}+a^2}\Big ]^{\frac{1}{2}}-1\Big )\ .
\end{align}
This efficiency can be written in the form 
\begin{align}
\eta (V^{}_{\rm ME}) =\eta_{\rm rev}[\sqrt{1+\widetilde{\zeta}}-1]/[\sqrt{1+\widetilde{\zeta}}+1]\ ,
\label{maxeta}
\end{align}
with the new figure of merit of the E2T device,
\begin{align}
\widetilde{\zeta} = \zeta \frac{(1-a^2)^2/(
\zeta^{}_{{\max}}+a_{}^2)
}{\Big (\Big [1-(1-a^2)\zeta/(\zeta^{}_{\max}+a^2)\Big ]^{\frac{1}{2}}_{}-a\Big )^2
}\ . \label{newzeta}
\end{align}
We plot $\widetilde{\zeta}/\zeta$ as a function of $a$ in Fig. \ref{fig3}, which shows that  $\widetilde{\zeta}$ 
can significantly exceed 
$\zeta$ of the 2T electronic device. 
As for $\eta(V_{\rm MP})$,  
$\widetilde{\zeta}$ has a maximum at  a certain positive $a$, which grows and moves to smaller values of  $a$ as $\zeta$ increases.  
The inset displays the region around $\zeta=\widetilde{\zeta}$. As seen, $\widetilde{\zeta}$ increasingly exceeds $\zeta$, making the E2T device better than the electronic 2T one.

In conclusion, we demonstrated that for our effective two-terminal quantum device, 
breaking the time-reversal symmetry yields efficiencies which can approach the Carnot efficiency, but are always lower than it.
It is  not sufficient to consider only the effective 2$\times$2 Onsager matrix. Including 
the restrictions from the entropy production of the underlying three-terminal device one finds  that the Carnot efficiency cannot be reached with a nonvanishing power. The efficiency of our device can exceed that in the 3T all-electronic device \cite{Brandner}.  We conjecture that similar restrictions apply  to other effective two-terminal devices, but a general proof 
remains a challenge for future work. An experimental realization of our model could  test the predictions concerning the advantages of phonon heat source  over an electronic one for producing electricity, in particular under the effect of a magnetic field.

\begin{acknowledgments}
Useful comments  given by G. Benenti are gratefully acknowledged. O.~E.~W.~and A.~A.~acknowledge support from  the infrastructure program of the Israel
Ministry of Science and Technology under Contract
No.~3-11173 and the kind hospitality of
the Institute of Industrial Science at the University of Tokyo.
K.~Y.~is supported by the Advanced Leading Graduate Course for Photon Science (ALPS), the University of Tokyo as well as by a Grant-in-Aid for Japan Society for the Promotion of Science (JSPS) Fellows (Grant No.16J11542). N.~H.~is supported by Kakenhi Grants No. 15K05200, No. 15K05207, and No. 26400409 from the Japan Society for the Promotion of Science.
\end{acknowledgments}


\begin{thebibliography}{99}
\bibitem{CommentB}
The magnetic field in the model of Ref. \onlinecite{PRB12} enters the Onsager matrix as a trigonometric function of the flux $\Phi$; strictly speaking, the entropy production can then vanish whenever $\sin\Phi$ vanishes. To avoid this complexity, the discussion is confined to small fields.
\bibitem{Commentin}
Note that since the junction comprises a thermal terminal that gives rise  inelastic scattering processes, the B\"{u}ttiker-Landauer scattering-matrix formalism used to study all-electronic  junctions \cite{Saito,Brandner,Balachandran,Brandner1} is not applicable for our model.
\end{thebibliography}
\end{document}